\documentclass[reprint,amsmath,amssymb,aps,prl,groupedaddress,nofootinbib]{revtex4-1}
\usepackage{graphicx}
\usepackage{amsthm,amssymb,amsmath,braket,mathdots}
\usepackage{bm}
\usepackage[hidelinks,pagebackref=false,pdfnewwindow=true]{hyperref} 
\usepackage{epstopdf,psfrag}
\usepackage{relsize,amsbsy}
\usepackage[export]{adjustbox}

\usepackage{graphicx,xcolor,tikz}

\newcommand{\be}{\begin{equation}}
\newcommand{\ee}{\end{equation}}

\newcommand{\bit}{\begin{enumerate}}
	\newcommand{\eit}{\end{enumerate}}

\definecolor{bananayellow}{rgb}{1.0, 0.88, 0.21}
\definecolor{straw}{rgb}{0.32, 0.28, 0.1}

\begin{document}

	\title{The Anomalous de Haas-van Alphen Effect in InAs/GaSb quantum wells}
		\author{Johannes Knolle}
                 \author{Nigel R. Cooper}
		\affiliation{\small T.C.M. Group, Cavendish Laboratory, JJ Thomson Avenue, Cambridge CB3 0HE, U.K.}
		
		\date{\today}

		\begin{abstract}
		 	  The de Haas-van Alphen effect (dHvAe) describes the periodic oscillation of the magnetisation in a material as a function of inverse applied magnetic field. It forms the basis of a well established procedure for measuring Fermi surface properties and its observation is typically taken as a direct signature of a  system being metallic. However, certain insulators can show similar oscillations of the magnetisation  from quantisation of the energies of electron states in filled bands. Recently the theory of such an {\it anomalous} dHvAe (AdHvAe) has been worked out but so far there is no clear  experimental observation. Here, we show that  the inverted narrow gap regime of InAs/GaSb quantum wells is an ideal platform for the observation of the AdHvAe. From our microscopic calculations we make quantitative predictions for the relevant magnetic field and temperature regimes, and describe unambiguous experimental signatures.      
		\end{abstract}

		\maketitle

{\it Introduction.}
In his seminal paper on the {\it Diamagnetism of Metals} in 1930~\cite{Landau1930} Lev Landau discovered what are now known as {\it quantum oscillations} (QOs) which describe the periodic variation of experimental observables as a function of applied magnetic field $B$. However, unaware of the experimental discovery in the same year of QO in the magnetisation~\cite{deHaas1930} -- the dHvAe -- as well as the conductivity~\cite{Shubnikov1930} -- the Shubnikov-de Haas effect (SdHe) --  Landau erroneously dismissed the effect to be unobservable small. The subsequent work of Onsager~\cite{Onsager1952} and Lifshitz-Kosevich (LK)~\cite{Lifshitz1958} showed the direct connection of the period of QOs to extremal areas of Fermi surfaces (FSs) and of the temperature dependence of the amplitude to the effective mass of electrons. This  quantitative theory has turned the effect into the most precise experimental tool for measuring FS properties~\cite{Shoenberg_1984} now standardly used around the world.

The dHvAe/SdHe originates from the periodic crossing of quantised Landau levels (LLs) through the chemical potential, such that measuring these effects in a material is almost taken as a synonym for the system being metallic~\cite{Shoenberg_1984}. Therefore, the observation of the dHvAe in the insulating state of the heavy-fermion material SmB$_6$~\cite{Tan2015} came as a big surprise. Motivated by this experiment, we have shown recently that, contrary to common intuition, QOs can appear in certain band insulators~\cite{Knolle2015} -- dubbed the {\it anomalous} dHvAe (AdHvAe). The prerequisites are that the filled and empty bands should be separated by a hybridisation gap that is on the order of the relevant cyclotron frequency, $\hbar \omega_c$, and that this hybridised region should trace out a well-defined closed surface in momentum space (a `shadow FS') at which the dispersion of the filled band changes abruptly. (In the cases described below this will be at the maximum of the filled band $E_{\rm max}$.) For a given magnetic field $B$ this results in a  LL structure with a sharp change in its dispersion as a function of LL index $n$ at the energy $E_{\rm max}$. The distance between LLs set by $\hbar \omega_c \propto B$ changes as a function of field such that subsequent levels are pushed over $E_{\rm max}$~\cite{FaWang2016} causing the thermodynamic potential to oscillate even in an insulator. 
\begin{figure}
	\centering
	\includegraphics[width=0.8\linewidth]{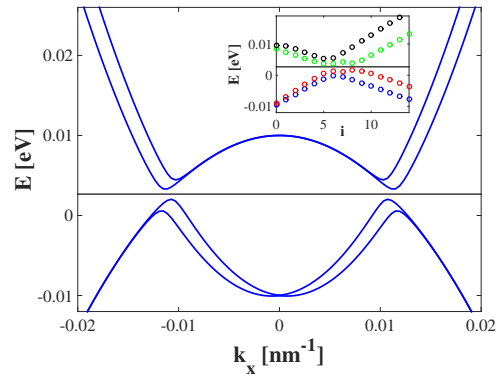}
		\caption{The  low energy band structure of InAs/GaSb QWs are shown for system close to the gap-closing transition with a small gap of $\Delta=1.3$meV. We assume that the chemical potential is in the middle of the gap (solid line). The inset shows the calculated Landau level structure for $B=0.6$T with up- and down-ward dispersing LL branches, as a function of an index $i=0,1,2\dots$ labeling the states in each band.}
	\label{Fig1}
\end{figure}

Our previous work~\cite{Knolle2015} was based on a flat band of f-electrons hybridised with a dispersive band of d-electrons. The connection of such a simple non-interacting toy model to the complicated Kondo insulator SmB$_6$ is currently under debate~\cite{Erten2016,Thomson2016,Xu2016,Knolle2016b}. However, subsequently the AdHvAe was shown to occur in other insulating band structures fulfilling the above mentioned prerequisites. Model calculations for topological insulators~\cite{FaWang2016} and gapped semi-metals~\cite{Pal2016} have explored the effect in more general settings, but an unambiguous experimental observation of the ultimately elementary AdHvAe is  missing.  

Here, we address the following outstanding question: What is a realistic and readily available experimental system for observing the AdHvAe? Is it possible to make quantitative predictions? We propose that the insulating regime of an electron hole bilayer with a small hybridisation gap is the system of choice. Specifically, InAs/GaSb QWs~\cite{Liu2008} constitute an ideal platform because they combine a number of desirable features: i) They have an inverted band structure resulting in up- and down-ward dispersing LLs, see Fig.\ref{Fig1} whose intersection provides a well-defined closed contour in momentum space; ii) Their dispersion and the band gap are highly tuneable as confirmed by the recent observation of the metal insulator transition~\cite{Qu2016,Karalic2016}; iii) They are well described by effective non-interacting models~\cite{Liu2008}; iv) They have small band masses resulting in sizeable cyclotron frequencies for small magnetic fields such that the AdHvAe should be observable in a broad regime of band gaps; v)    
 In contrast to HgTe QWs~\cite{Bernevig2006}, which are very difficult to fabricate~\cite{Konig2007}, InAs/GaSb QWs are much simpler with many different groups studying the quantum spin Hall properties~\cite{Du2015,Muraki2016,Papaj2016,Mu2016,Dyer2016,Qu2016,Mueller2015}.    

By performing a first quantitative calculation we show that the AdHvAe is straightforwardly accessible in InAs/GaSb QWs for magnetic fields below 2 Tesla. A direct observation of QO in the magnetisation and the simultaneous absence of QO in the conductivity (no SdHe) will be a smoking gun signature of its discovery.

{\it Effective 4-band model.}
 The low energy electronic degrees of freedom can be described by an effective four band model in the basis of electron/hole-like (e/h) states  $[e \uparrow,h \uparrow, e\downarrow,h\downarrow]$~\cite{Liu2008}. The Hamiltonian, which is similar to the BHZ model describing the quantum spin Hall effect in HgTe/CdTe QWs, is described by~\cite{Liu2008,bookFranz}
\begin{widetext}
\begin{eqnarray}
\label{Hamp}
\hat H =  
\begin{pmatrix}
M_0+\mu_+ k^2 - \frac{g_e \mu_B}{2} B& \beta k_+ & \Delta_+ k_+ - i \alpha k_- & -\Delta_0\\
\beta k_- &  -M_0-\mu_- k^2 - \frac{g_h \mu_B}{2} B & \Delta_0 & \Delta_- k_- \\
\Delta_+ k_- + i \alpha k_+ & \Delta_0 & M_0 + \mu_+ k^2 +\frac{g_e \mu_B}{2} B & - \beta k_-\\
-\Delta_0 & \Delta_- k_+ & -\beta k_+ & - M_0 - \mu_- k^2 + \frac{g_h \mu_B}{2} B
\end{pmatrix} 
\end{eqnarray}
\end{widetext}
with $k_+=(k_x+i k_y)$, $k_-=k_+^*$, $k^2=k_x^2+k_y^2$. The parameter $\beta$  controls the degree of hybridisation between the sub-band systems, so largely determines the size of the hybridisation gap. In actual materials it can be widely controlled, by varying for example different layer thickness or backgating~\cite{Liu2010,Du2015,Qu2016}. 

The terms proportional to $\Delta_i$ ($i=+,-,0$)describe the bulk inversion asymmetry (BIA) and $\alpha$-terms arise from the structural inversion symmetry (SIA). In contrast to HgTe/CdTe QWs the electron and hole subbands are localised in spatially separated layers in InAs/GaSb QWs. Therefore, the inversion symmetry is broken and the SIA terms dominate over the BIA terms. Since the latter are at least an order of magnitude smaller we can safely neglect them ($\Delta_i=0$ in the following) which facilitates the calculation of the quantised LL dispersion in an orbital magnetic field. Finally, we have included a Zeeman term which will lift some of the remaining spin degeneracies. Typical values have been determined very recently, and are around $g_e\approx 10$ and $g_h\approx 3$\cite{Mu2016}. In Fig.\ref{Fig1} we show the band structure for a QW structure close to the critical thickness with a small band gap of $1.3$ meV. The microscopic parameters have been calculated before and we use the same values as Ref.~\cite{Pikulin2014} taken from Ref.~\cite{bookFranz} except for a smaller $\beta=0.12$[eV\AA] which reduces the gap bringing the system closer to the metal-insulator transition.

In the following we will set up the calculation for the energy levels in an orbital magnetic field~\cite{Markus2008,Liu2010,Zhang2014} which can then be used to directly calculate experimental observables.  We introduce $B$ in the out of plane direction via the vector potential $\mathbf{A}$ which is minimally coupled to the crystal momentum  such that $\mathbf{\Pi}=\mathbf{k}-\frac{e}{c} \mathbf{A}$. Then we can replace the momentum operators in Eq.\ref{Hamp} with the standard ladder operators $k_+ \to \frac{\sqrt{2}\hbar}{l_B} \hat a^{\dagger}$,  $k_- \to \frac{\sqrt{2}\hbar}{l_B} \hat a$ and $k^2 \to \frac{2\hbar^2}{l_B^2}( \hat a^\dagger \hat a+ \frac{1}{2})$ with the magnetic length $l_B=\sqrt{\frac{\hbar c}{e |B|}}\approx \frac{26 \text{nm}}{\sqrt{B [\text{Tesla}]}}$. With the Ansatz wave function $| \Psi_n \rangle = \left[u_n |n  \rangle, v_n |n-1  \rangle, w_n |n+1  \rangle, x_n |n+2  \rangle \right]^T$ in terms of the standard harmonic oscillator states $|n_o\rangle$ the problem of calculating the energy levels, $\hat H | \Psi_n \rangle = E_n | \Psi_n \rangle$,  for a given field $B$ as a function of LL index $n$ is reduced to that of finding the eigenvalues of
\begin{widetext}
\begin{eqnarray}
\label{Hamn}
 H_n =  \!
\begin{pmatrix}
M_0\!+\!\frac{2 \mu_+}{l_B^2} (n\!+\!\frac{1}{2}) \!-\! \frac{g_e \mu_B}{2} B& \frac{\sqrt{2}\beta}{l_B} \sqrt{n} & - i \frac{\sqrt{2}\alpha}{l_B} \sqrt{n\!+\!1} & 0\\
\frac{\sqrt{2}\beta}{l_B} \sqrt{n} &  -M_0\!-\!\frac{2 \mu_-}{l_B^2} (n\!-\!\frac{1}{2}) \!-\! \frac{g_h \mu_B}{2} B & 0 & 0 \\
 i \frac{\sqrt{2}\alpha}{l_B} \sqrt{n\!+\!1} & 0 & M_0 \!+\! \frac{2 \mu_+}{l_B^2} (n\!+\!\frac{3}{2}) \!+\!\frac{g_e \mu_B}{2} B & - \frac{\sqrt{2}\beta}{l_B} \sqrt{n\!+\!2}\\
0 & 0 & - \frac{\sqrt{2}\beta}{l_B} \sqrt{n\!+\!2} & - M_0 \!-\! \frac{2 \mu_-}{l_B^2} (n\!+\!\frac{5}{2})\!+\! \frac{g_h \mu_B}{2} B
\end{pmatrix}. \nonumber
\end{eqnarray}
\end{widetext}
[Note that the full matrix, coupling all four bands, only applies for $n>0$; it reduces to $3+n$ coupled bands for $n=0,-1,-2$ since the entries Ansatz must have oscillator index $n_o\geq 0$.]
In the inset of Fig.\ref{Fig1} we show the evolution of the LL branches for a field of $0.6$ Tesla. For increasing LL index $i$ (labelling the states in each band) the lower branches first disperse upwards before they turn downwards after reaching the hybridised gapped region around $i=7$. For increasing the magnetic field, the distance between the levels increases such that each of the levels will be pushed consecutively over the maximum of the LL branch -- a prerequisite for the AdHvAe.

With the (numerically) calculated energy levels $E_n^\alpha$ (with
$\alpha$ labelling the energies of each matrix $H_n$) one can
directly calculate the magnetisation from the grand potential
via
\begin{eqnarray}
M=-\frac{\partial}{\partial B} \Omega \  \text{with} \  \Omega= -k_B T N_{\Phi}\sum_{n,\alpha} \ln \left[ 1+e^{\frac{\mu-E^{\alpha}_n}{k_B T}}\right]
\end{eqnarray}
with the Landau level degeneracy $N_{\Phi}=\frac{BA}{\Phi_0}$ (here
$A$ is the area of the 2D system and $\Phi_0=\frac{hc}{e}$ is the flux
quantum). However, since 
 the energies of the two lower LL branches are unbounded
from below for increasing $n$, the sum
 is divergent at its upper limit $n\to\infty$. This is, of course, unphysical and an artefact
of the continuum approximation. It can be easily cured by subtracting
the grand potential for a simple band insulator, obtained for
non-inverted levels with $M_0 \to \infty$, for which the only occupied levels are $\tilde{E}^1_n= - \frac{2 \mu_-}{l_B^2} (n+\frac{5}{2})$ and $\tilde{E}^2_n= -\frac{2 \mu_-}{l_B^2} (n-\frac{1}{2})$.  This simple band insulator has a similarly
divergent potential within the continuum limit, but a net magnetization that must vanish when effects beyond that limit are included.
Therefore, we can obtain the regularized magnetization $M$ for our model from the difference
\begin{eqnarray}
\frac{\tilde \Omega}{k_B T N_{\Phi} }=-\!\sum_{n,\alpha}  \ln \left[ 1+e^{\frac{\mu-E^{\alpha}_n}{k_B T}}\right] \!+ \!\!\sum_{n, \alpha=1,2}  \ln \left[ 1+e^{\frac{\mu-\tilde E^{ \alpha}_n}{k_B T}}\right]. \nonumber
\end{eqnarray}

A simple but experimentally crucial observation is the following: only thermodynamic observables, e.g. the magnetisation or its susceptibility, oscillate as a function of field because they are determined by the sum over all occupied energies~\cite{Knolle2015}.  However, other observables like charge transport or NMR relaxation rates do not oscillate because they are  determined by the DOS at the chemical potential which is, of course, zero in the insulating regime~\cite{Pal2016}.  These simply have a vanishing contribution for temperature scales below the activation gap~\cite{FaWang2016}. 

Because the calculation of the charge conductivity requires a precise knowledge of scattering mechanisms from defects and interactions, here, we only calculate a proxy of it. The main qualitative behaviour of transport properties is captured by the evolution of the temperature dependent density of states at the Fermi level ($\mu$DOS)~\cite{FaWang2016} 
\begin{eqnarray}
\mu\text{DOS }\! = \sum_{n,\alpha}   \frac{\partial n_F (E_n^{\alpha})}{\partial \mu} =
\frac{1}{2k_B T} \sum_{n,\alpha} \frac{1}{\cosh \left[ \frac{E_n^{\alpha}-\mu}{k_B T}\right]+1} \nonumber
\end{eqnarray}
with the standard Fermi function $n_F$.

{\it Results.}
\begin{figure}
	\centering
	\includegraphics[width=1.0\linewidth]{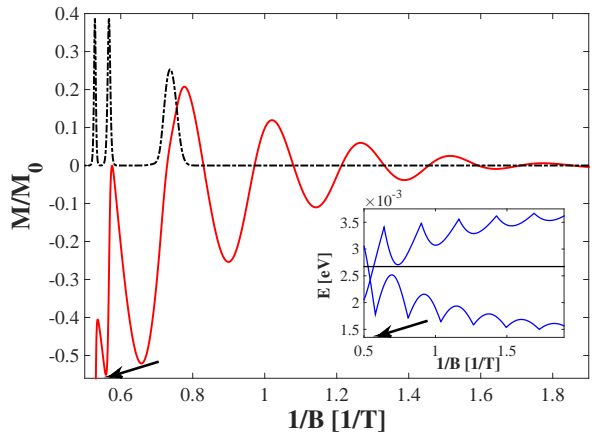}
		\caption{QO as a function of inverse magnetic field in the narrow gap insulating regime with the same parameters as used in Fig.\ref{Fig1} and for $T=0.3$K. The amplitude of the magnetisation $M$ (red solid) is scaled by the QO amplitude $M_0$ in the unhybridised electron-like metallic subsystem, see text. The evolution of the $\mu$DOS, which is expected to follow the qualitative behaviour of charge transport, is also shown (black dashed line). Note the magnetic field induced gap closing (black arrow) which can also be traced by the evolution of the extremal energy levels around the chemical potential shown in the inset.}
	\label{Fig2}
\end{figure}	
In the following we focus on the small-gap insulating regime of InAs/GaSb QWs which is expected to be quantitatively described by the band structure shown in Fig.\ref{Fig1}. The amplitude of the AdHvAe is only observable in the regime where the hybridisation gap, $\Delta$, is not considerably larger than the energy scale associated with the LL quantisation of the energies~\cite{Knolle2015} otherwise it is exponentially suppressed. This scale is given by the cyclotron frequency $\hbar \omega_c = \frac{2 \mu_+}{l_B^2}$ and determines the magnetic field range in which the AdHvAe is observable. For InAs/GaSb QWs it varies from about $3$meV for $B=1.7$T to $1$meV at $B=0.6$T which is easily reachable experimentally. 

\begin{figure}
	\centering
	\includegraphics[width=1.0\linewidth]{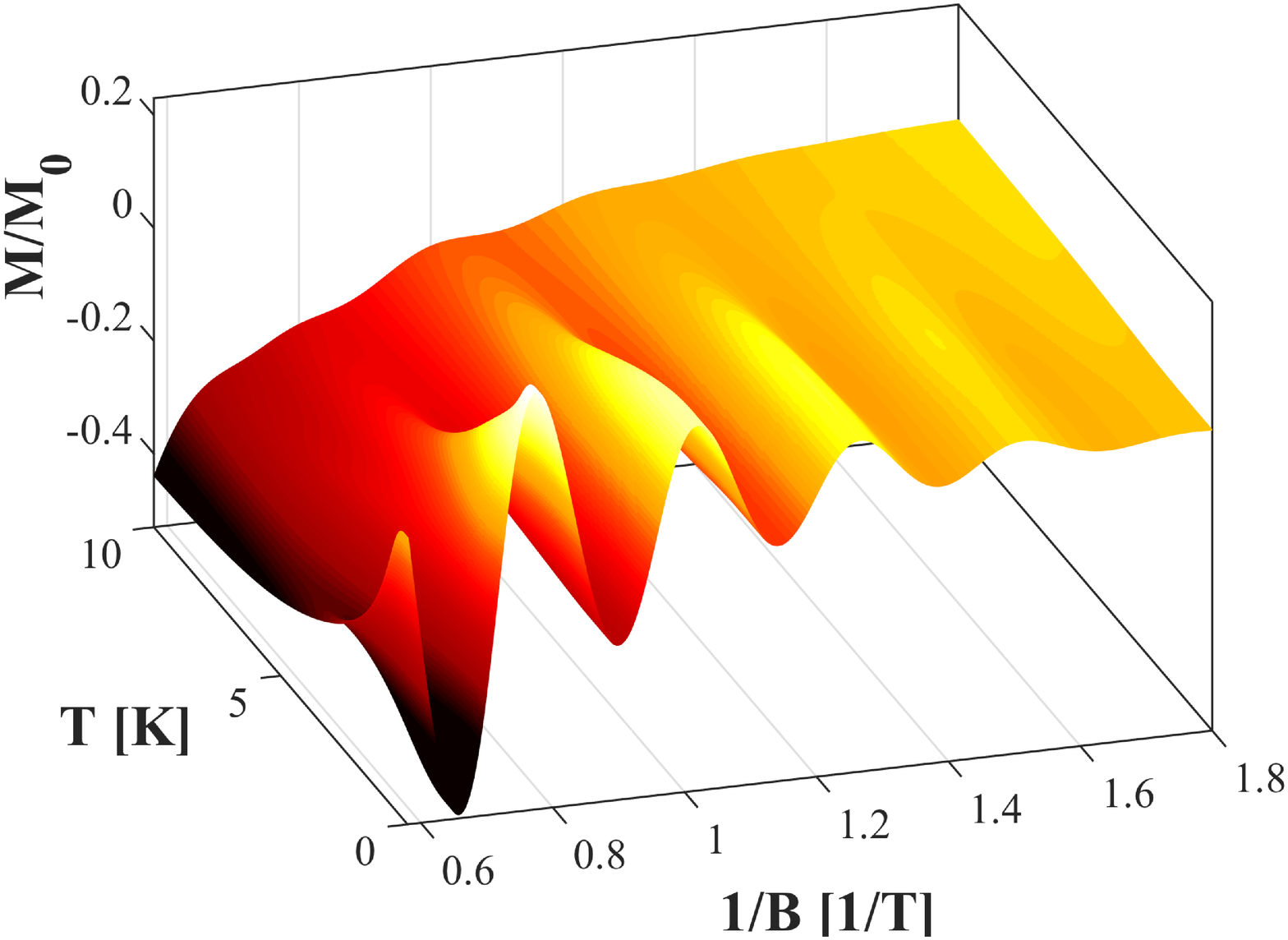}
	\includegraphics[width=1.0\linewidth]{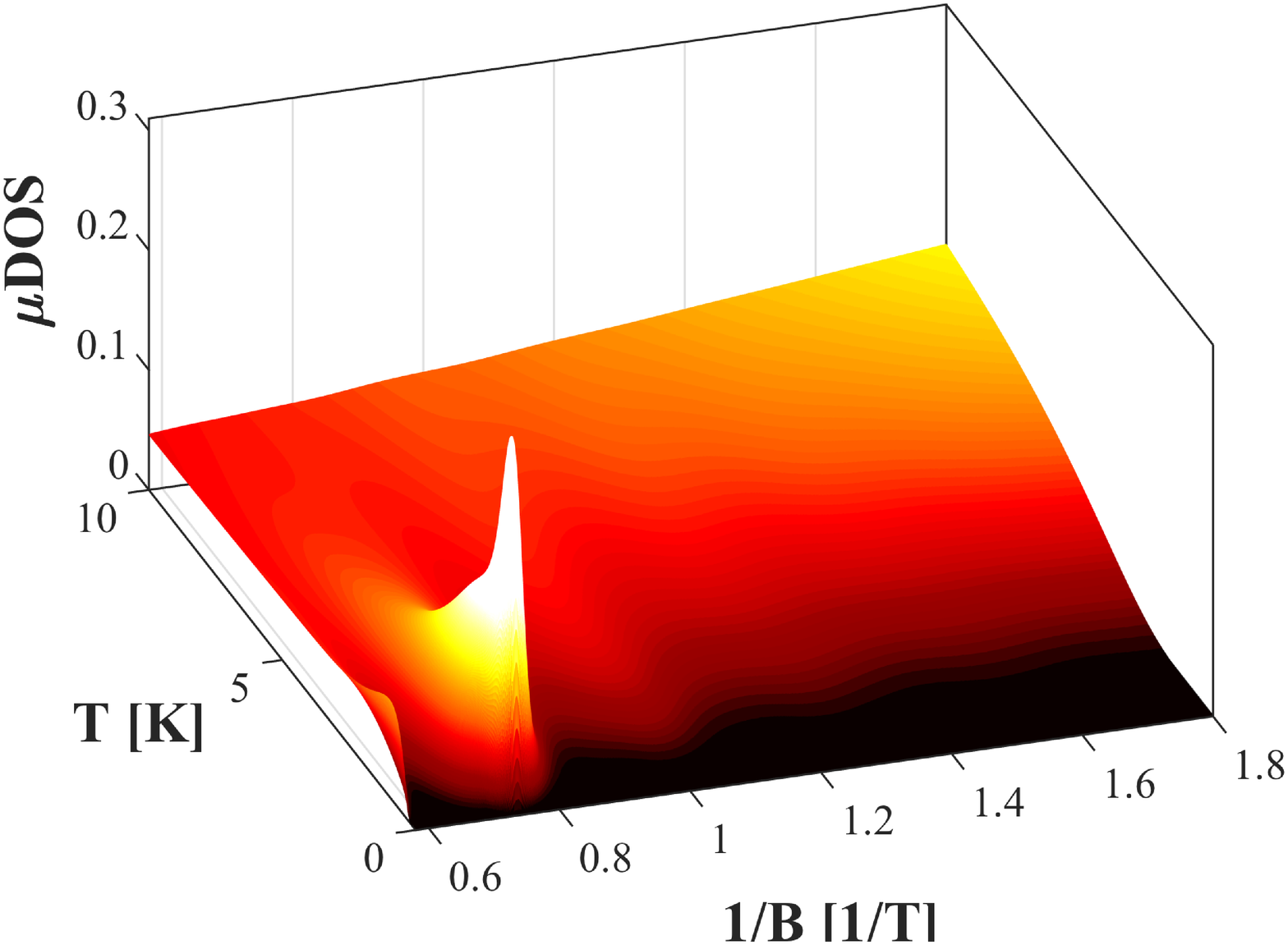}
		\caption{The evolution of the magnetisation (upper panel) and the $\mu$DOS (lower panel), which mimics the behaviour of charge transport, as a function of temperature.}
	\label{Fig3}
\end{figure}	
We use the same parameters as before  and show in Fig.\ref{Fig2} the magnetisation (red solid curve) as a function of inverse magnetic field for a temperature $T=0.3$K. The inset shows the evolution of the extremal energy levels closest to the chemical potential highlighting the fact that for inverse fields below $1/B\approx 0.57/\mbox{T}$ (black arrow) the gap closes and the system becomes metallic~\cite{FaWang2016}. At this point the amplitude of the magnetisation increases. However, the central result of this work is that clear QOs of the magnetisation persist  even in the insulating regime! The frequency is given by the area of the gapped `shadow FS'. Note, the amplitude of the oscillating magnetisation $M$ is expressed in units of the typical amplitude $M_0=\frac{\hbar \omega_c \rho A}{\pi B}$~\cite{Shoenberg_1984} for oscillations of the electron like band in the unhybridised metallic regime (here $\rho=\frac{A_k}{2 \pi^2}$ is the density set by the relative FS area $A_k$). The amplitude of the QOs in the insulator is about an order of magnitude smaller than in the metal, but should still be detectable in experiments.

In addition, we show in Fig.\ref{Fig2} the evolution of the $\mu$DOS (black dashed) which mimics the behaviour of charge transport. Note we have checked that for chemical potentials inside the bands the QO period directly scales with the FS area as expected for the standard SdHe and dHvAe. Here, sharp peaks appear once the system becomes metallic below $1/B\approx 0.57/\mbox{T}$ (black arrow) but in contrast to the magnetisation there are no oscillations in the insulating regime! The smaller broad peak appearing around $1/B\approx 0.73/\mbox{T}$ is related to the small gap, see inset, on the order of $k_B T$ which leads to thermally excited states at the Fermi level but it also disappears for lower $T$.   

In Fig.\ref{Fig3} upper panel we show the temperature evolution of $M/M_0$. Note the small drift as a function of temperature and field which is related to the fact that two bands with different gap values contribute. Moreover, the gap values are dependent on the magnetic field and the complex behaviour of the LL dispersion can even result in temperature induced phase jumps of the oscillations~\cite{FaWang2016}.  

Both the amplitude of $M$ and $\mu$DOS deviate substantially from the standard LK behaviour. In standard metallic systems the oscillation amplitude is always monotonously decreasing as a function of $T$~\cite{Shoenberg_1984} whereas in the AdHvAe it has plateau at lowest temperatures or even a maximum at a temperature $T^*$, which is set by the distance of the LL extrema to the chemical potential~\cite{Knolle2015}. In the future this could be used to measure the size of the insulating gap $\Delta$ and the position of $\mu$. In the upper panel of Fig.\ref{Fig3} the maximum of the amplitude moves to higher temperatures for increasing $1/B$ which traces the evolution of the gap, see inset in Fig.\ref{Fig2}.  
Further increasing the temperature the amplitudes decrease until they are completely washed out around 10K (above $k_BT\approx \hbar \omega_c$).

The lower panel of Fig.\ref{Fig3} displays the evolution of the $\mu$DOS. While there are no oscillations at $T=0$ in the insulating regime they appear at nonzero temperatures from thermally excited states. Nevertheless they are very weak except for the peak at $1/B\approx 0.73/$T associated with a very small gap, see inset in Fig.\ref{Fig2}. The weak oscillations  quickly disappear with increasing temperature.

Finally,  we have checked that the behaviour is qualitatively similar for different values of the Zeeman coupling (g-factors) or gap values with only slight changes from a change of the LL gap and the critical field above which the system becomes gapless. For example, for a much bigger gap of $\Delta=6.3$meV ($\beta=0.36$) we find that there are still appreciable QOs in $M$ for fields above $3$T (not shown). Overall, the AdHvAe is quite robust as long as the lower LL branches disperse up- and down-wards and $\Delta$ does not greatly exceed $\hbar \omega_c$. Hence, also the inclusion of the BIA terms or other small perturbations will only induce small quantitative changes.

{\it Discussion.}
We have shown that the AdHvAe is observable in the narrow gap insulating regime of InAs/GaSb QWs. For reasonable band gaps $\approx 1.5$meV magnetic oscillations appear at temperatures below about 5 Kelvin for magnetic fields around 1 Tesla, with an amplitude that is roughly one order of magnitude smaller than for the corresponding gapless electron sub-system. Hence, the experimental constraints are well within reach and the remaining experimental challenge should be the measurement of the magnetisation of InAs/GaSb QWs, e.g. via magnetic torque as has been carried out for other 2D electron gases~\cite{Schwarz2000,Harris2001,Schwarz2002}. The predicted behaviour contrasts strongly to standard QOs in metals, where both the SdHe and the dHvAe occur simultaneously. For the AdHvAe in insulators only observables related to the thermodynamic potential oscillate whereas those quantities determined by the DOS at the chemical potential, e.g. charge transport, vanish in the low temperature limit. Hence, the observation of QOs in the magnetisation without a SdHe will be an unambiguous signature of the AdHvAe.   Finally we note that the amplitude of QOs depends strongly on the hybridisation gap and on the chemical potential (at nonzero $T$), so this effect could provide a powerful tool for measuring gap sizes and the position of the chemical potential in narrow gap insulators.

\section*{Acknowledgements}
We acknowledge helpful discussions with Kiryl Pakrouski and especially for bringing Refs.~\cite{Qu2016,Karalic2016} to our attention. The work is supported by a Fellowship within the Postdoc-Program of the German Academic Exchange Service (DAAD) and by the EPSRC Grant No. EP/J017639/1.
Statement of compliance with EPSRC policy framework on research data: All data accompanying this publication are directly available within the publication.

\bibliography{QO_refs}

\end{document}